\documentstyle[12pt, aasms4]{article}
\begin{document}

\title{Two Groups of Nearly Coeval Star Clusters 
in the Small Magellanic Cloud\footnotemark}

\author{R. Michael Rich}
\affil{Department of Physics and Astronomy, University of California
at Los Angeles}
\affil{8979 Math-Sciences Bldg, Los Angeles, CA 90095-1562\ \  rmr@astro.ucla.edu}
\author{Michael Shara}
\affil{Space Telescope Science Institute}
\affil{3700 San Martin Drive, Baltimore MD, 21218\ \ shara@stsci.edu}
\author{S. Michael Fall}
\affil{Space Telescope Science Institute}
\affil{3700 San Martin Drive, Baltimore MD, 21218\ \ fall@stsci.edu}
\and
\author{David Zurek}
\affil{Space Telescope Science Institute}
\affil{3700 San Martin Drive, Baltimore MD, 21218\ \ zurek@stsci.edu}

\footnotetext{
Based on observations with the NASA/ESA Hubble Space Telescope,
obtained at the Space Telescope Science Institute, which is 
operated by the Association of Universities for Research 
in Astronomy, Inc., under NASA contract No. NAS5-2655.}

\begin{abstract}

We report new photometry of populous intermediate-age
clusters in the SMC using the {\sl Hubble Space Telescope}.
In contrast to the accepted picture, these clusters appear to
have formed in two brief 
intervals, one $8\pm 2$ Gyr ago, and a more recent
burst $2\pm 0.5$ Gyr ago.  When the ridgelines of the 
four clusters (NGC 339, 361, 416, and Kron 3)
in the 8 Gyr burst are aligned, the
dispersion in turnoff luminosities is $<0.2$ mag,
corresponding to a maximum age spread of $\pm 0.7$ Gyr.
When the ridgelines of three clusters (NGC 152, 411, and
419) in the 2 Gyr burst are
aligned, the maximum dispersion of 0.2 mag in
turnoff luminosity corresponds to 
a permitted age spread of $\pm 0.2$ Gyr.
Within each group of clusters, the entire cluster loci (including
red giant branches and clumps) are nearly identical, consistent
with indistinguishable metallicities and ages.
In contrast to the wide dispersion in ages previously reported
in the literature, our sample with more precise photometry and
age measurements
supports a burst-punctuated rather than a continuous cluster formation
history for the 2 Gyr and 8 Gyr SMC clusters.

\end{abstract}


\section{Introduction}

The age distribution of star clusters in the Large Magellanic Cloud (LMC)
has long been known to be approximately bimodal
(van den Bergh 1991; vdB91, Westerlund 1997), while 
the Small Magellanic Cloud (SMC) is considered to have formed
its clusters continuously over the last 12 Gyr (Olszewski, Suntzeff,
Mateo (1993; OSM93, vdB91).  The ages of the Magellanic 
clusters have been
difficult to measure because of stellar crowding, field star
contamination, and faintness of the main sequence turnoffs. 
The availability of large aperture telescopes and electronic
detectors have vastly increased the number of Magellanic
clusters with age determinations.  In the age range from 1-14 Gyr,
where the cluster ages potentially preserve the fossil record of
ancient cluster formation and disruption history, 
the LMC has a period from 2.5-12 Gyr in which
only one star cluster (ESO121-SCO3) 
appears to have been formed (OSM93).  In contrast,
existing data suggest that the SMC was 
more or less continuously forming clusters
during this interval (Mighell et al. 1998).  Furthermore, the 
oldest clusters in the LMC appear to be approximately as old
as the oldest globular clusters in the Galaxy (Mighell et al. 1996,
Olsen et al. 1998, Johnson et al. 1999)
while the oldest cluster in the SMC is distinctly
younger (Stryker et al. 1985, Shara et al. 1998).  
It has been argued (vdB91)
that the disparate timing
of these bursts is inconsistent with scenarios in which cluster
formation was triggered by encounters between the SMC and LMC.

Much of our understanding of the evolution of the Magellanic Cloud/Galaxy
system is limited by the depth and accuracy of ground-based
photometry.  Further, the derived cluster ages are not consistent,
both due to the use of different telescopes, filters, and detectors,
and due to the different isochrones used to 
fit the main sequence turnoff.
Finally, ground-based color-magnitude diagrams must avoid the cluster
cores because of crowding and thus suffer serious contamination by
field stars.

In order to obtain a large, uniform sample of Magellanic Cloud
cluster photometry, 
we have undertaken a survey of short-exposure images in
the cluster cores using the
HST.  The results on nearly 50 clusters will be described elsewhere (Rich
et al. 1999).
Taking advantage of this uniquely homogeneous and high quality data set,
we focus here on our sample of 7 SMC clusters with ages $>1$ Gyr,
in which we find evidence for two remarkable bursts.
Our sample of SMC clusters includes all clusters with
$V<13$ and $(B-V)>0.5$ in Table II of van den Bergh's (1981) compilation.
We therefore determine ages for the most luminous $(M_V<-6)$ old
SMC clusters -- those which most closely resemble their Milky Way
counterparts.

After producing the color-magnitude diagrams we noticed that our
sample naturally divides into two groups of clusters within which there
is very little dispersion in age or metallicity.  The following
section describes our photometry and method for comparing the
cluster color-magnitude loci.  Section 3 describes our age determination
for the 2 and 8 Gyr cluster groups using the most straightforward
method of measurement, the magnitude
difference between the horizontal branch and the turnoff
point and the equivalent approach using
isochrone fitting.  In this section, we also consider
the minimum possible age dispersion among these clusters.
 Section 4 considers the maximum age
dispersion that could be derived from our data.  We consider our new finding
relative to what is already known about the oldest SMC clusters in
Section 5, and we report our conclusions in Section 6.

\section{Photometry and Color-magnitude Diagrams}

The SMC clusters were observed as part of a snapshot survey using
WFPC2 on board the {\sl Hubble Space Telescope}.  Table 1 gives the log of 
observations; all exposures were shorter than 600 sec, and were obtained
through the  F450W and F555W filters (one frame per color).  The frames were 
reduced according to the HST pipeline; however we have used the best 
calibrations available rather than accept the data as processed near the time 
of acquisition.

Magnitudes were measured from the images using the point-spread function (PSF) 
fitting program ALLSTAR in the DAOPHOT II environment (Stetson 1994a,b).  The PSF's
were created for each cluster in each color by using the standard approach of 
subtracting neighboring stars from every star contributing to the PSF, a 
procedure that is repeated until it is judged that there is no improvement 
between iterations.  For the first cut aperture photometry, we use a 3 pixel 
aperture.  The final photometry must be corrected to a large aperture of 0.5''
(or 11 pixels on the PC).  The aperture correction is measured for PSF stars 
from which the near neighbors have been subtracted.  This produces an aperture 
correction for each image (Table 2) which varies due to small changes in the HST
focus.  Because most of the observations were taken before the cool down of the 
WFPC2 on 23 April 1994, a correction (Table 3) for the throughput difference is 
made (HST Data Handbook: WFPC2 Photometric Corrections).  To be complete, we 
have also included a small correction for the contamination that accumulates 
between warm-ups of the cameras; this is a very small effect for optical 
magnitudes
(Table 4).  Finally, the photometry was transformed
to $B$ and $V$ respectively as described in Holtzman et al. 1995b.

\subsection{Color-Magnitude Diagrams}

The color-magnitude diagrams resulting from the photometry naturally
divide into two groups based on the magnitude difference between
the red horizontal branch and the main sequence turnoff point.
For NGC 411, 152, and 419 this difference is about one magnitude,
while it is about 2.5 magnitudes for Kron 3, NGC 339, 361, and 416.
We overplot the ridgeline of NGC 411 in the younger group (which
we name the NGC 411 group) and we overplot that of Kron 3 relative to
the older ``Kron 3'' group of clusters.  Within each group of clusters,
it is clear that the color-magnitude arrays are remarkably similar.
It is especially noteworthy that the age-sensitive turnoff to horizontal
branch magnitude difference, and the metallicity sensitive giant branch
position, is very similar with each group.  While the similarity
is evident in the Kron 3 group, the ridgeline does not overplot those
data perfectly.  However, if the Kron 3 ridgeline is shifted by eye,
the agreement with each of the other older clusters is excellent; 
we thus speculate that the dispersion in the apparent clump luminosity
could be due to a dispersion in distance, extinction, or calibration
uncertainty.  To study further the possible similarity of clusters
within each group, we fit empirical ridgelines by eye to each cluster
and shift them freely to coincide with the NGC 411 ridgeline (younger
group) and Kron 3 ridgeline (older group) respectively.   The dispersion
in relative ages and metallicities can be compared within each group,
even though the distance modulus to any individual cluster is
relatively uncertain.

The empirical cluster ridgelines are the result of a freehand fit by eye to the 
color-magnitude diagrams.   The ridgelines are given in Tables 6-8.  
Within each group, these CMD loci are nearly identical.
We have exploited these similarities and shifted the loci 
in $V$ and $B-V$ to match Kron 3 in the case of the older clusters, and NGC 411 
for the younger clusters.  We report the shifts required for the clusters to 
coincide with the designated templates in Table 9, which also gives the mean 
luminosity of the red clump for these clusters.   Given the many corrections 
required to our data to establish the calibration, we consider our zero points 
to be accurate to $\approx 0.05$ mag.

It is interesting to note the small dispersion in red clump luminosities
for the clusters.  The old clusters have $V=19.46 \pm 0.19$ and
the young clusters have $V=19.50 \pm 0.08$, neither being statistically
different from the sample mean of $V=19.48 \pm 0.14$.  We find no
evidence that the luminosity of the red clump differs between these
groups of clusters, which would be expected from theoretical
calculations that find little dependence of clump luminosity on 
age or metallicity (cf. Sweigart, Greggio \& Renzini, 1990;
Castellani, Chieffi \& Straniero 1992; Girardi et al. in preparation).
The red clumps of our clusters are fainter than the 
clump for the SMC field of $V=19.25\pm 0.05$ derived from
the data of Udalski et al. (1998).  We note, however, that the
mean $V$ was calculated by adding an approximate $V-I$ for the
clump to the derived peak $I$ magnitudes of Udalski et al.

\section{Cluster Age Measurements}

The apparent shifts among the clusters in the Kron 3 group
are due either to differing distance modulus, reddening, or
calibration uncertainties.  However, we focus our efforts
on learning as much as possible from the similarity of the
color-magnitude arrays and loci within each group.  
The best constraint on relative ages is the $\Delta V^{HB}_{TO}$
index, the magnitude difference between the turnoff and the
horizontal branch.   The main sequence turnoff luminosity
fades with increasing age, while the HB luminosity is
determined by the mass of the helium burning core, which
is roughly constant in low-mass stars.  The age is derived
by using various calibrations of $\Delta V^{HB}_{TO}$ 
versus age and metallicity from
the literature (e.g. Walker, 1992).  The age may also be derived
by fitting the isochrones to the red clump rather than shifting
the isochrones by a prescribed distance modulus and reddening.
Although this method relies on essentially the same fundamental
physics, the isochrones are fit to the entire color magnitude
diagram and the result is less sensitive to the increase of
errors at faint magnitudes.  In our first approach to exploring
the age dispersion of the clusters, we shift the ridgelines to
agree as closely as possible with the template cluster and then
determine the maximum age dispersion among the clusters implied
by this procedure.  In Section 4, we will also consider the age dispersion
derived from independent fits to the apparent color-magnitude
arrays of each cluster.

\subsection{The 2 Gyr Old Burst}

We consider three intermediate-age clusters, which we designate the
NGC 411 group.  Figure 1 illustrates the unshifted NGC 411 ridgeline
overplotted on the data of NGC 411, 152, and 419. 
The cluster loci are identical to 
within observational uncertainty, after
the offsets in Table 9 are applied.  A
fit to the red clump using the Z=0.001 Bertelli et al. (1994)
isochrones (consistent with [Fe/H]=$-0.9$ from Da Costa \& Mould 1986)
gives $2.0\pm 0.2$ Gyr.  Because the fit of the red giant branch
is good, we conclude that the metallicity of the isochrones is
in good agreement with the data.  In order to explore the possible
spread in age within the 2 Gyr group, we must know the maximum
possible spread in metallicity between these clusters.
We estimate the permitted dispersion in metallicity from the rate
of color change in the red giant branch locus as a function of metallicity,
as measured from the isochrones from the isochrones.  Just below the
RGB tip, this is approximately $\Delta \rm [Fe/H] = 3.1 \Delta (B-V)_0$
and the entire spread in the 2 Gyr clusters corresponds to 
0.15 dex in metallicity.  The nearly perfect agreement of the full
cluster ridgelines (including the red giant branches) within this
group of clusters is remarkable, and admits very little variation in
age, metallicity, or any other significant parameter.

The age estimate from
$\Delta V^{HB}_{TO}$ (Walker 1992) gives $1.65< t_{Gyr} < 2.05$.  
The lowest age would apply for $\Delta V^{HB}_{TO}=1.1\ \rm mag$
and [Fe/H]=--1.1, while the largest age would correspond
to $\Delta V^{HB}_{TO}=1.3$ and [Fe/H]=--1.6.  In fact, the range
due to the difficulty of locating the
nearly vertical turnoff point in the noisy data rather than an actual
dispersion, and the metallicity differences within this group are
also almost certainly smaller than stated.  Our preferred estimate
from this method is an age of $1.85 \pm 0.1$ Gyr for the NGC 411
group.

The turnoff point of a young star cluster will fade more rapidly
with age than that of an older cluster, such as those in the Kron 3
group discussed below.
Buonanno et al. 1993 give $\Delta \log t_9 = (0.44+0.04[Fe/H]\Delta M_V(TO)$,
and the small dispersion in relative cluster loci constrains the
full age spread to be less than 0.4 Gyr.  If all of the cluster
ridgelines are forced to coincide and the isochrones are superposed,
then direct inspection of the result (Figure 1) suggests that the
full age dispersion could be as small as 0.2 Gyr, or 10\% of the
age of the clusters.
%

\subsection{The 8 Gyr Old Burst}

In Figure 2, we display the color-magnitude diagrams of the 
four SMC clusters in the older of the
two bursts, which we designate the ``Kron 3'' group.  
We have overplotted the empirical Kron 3 ridgeline on all four
CMDs, without applying any magnitude or color shifts. 
The nearly precise agreement of the Kron 3 ridgeline with the
ridgelines of NGC 339, 361, and 416 is striking.  The combination of
nearly identical main sequence turnoffs, red giant branches,
and clumps is consistent with a very small dispersion 
in age or metallicity within this group
of clusters.   In contrast to the other clusters in this group,
NGC 416 is superposed near the center of the SMC, and we speculate
that it may well be reddened relative to the other clusters (explaining its fainter and redder clump and locus).  We will return to the reddening
of the clusters in Section 4.  Although WFPC2 photometry of the oldest
SMC clusters has been reported in the literature before (Mighell et al. 1998)
our photometry of these clusters is permits us to
see clearly delineated the turnoff point, subgiant branch and 1-2 mag of main sequence.

Figure 2 also shows our best fit 
(isochrones force-fit at the red clump) 
using the Z=0.001 (corresponding to [Fe/H]=$-1.27$)
Bertelli et al. (1994) isochrones, which yields an age of $8.0\pm 0.5$ Gyr.
Recent Ca index abundance measurements by
Da Costa \& Hatzidimitriou (1998) give $-1.12$ dex for Kron 3 and
$-1.46$ for NGC 339; our isochrones are consistent with this range of metallicity.
Applying the age-sensitive magnitude difference between the turnoff and
HB, $\rm \Delta V^{HB}_{TO}$, we find $\Delta V^{HB}_{TO}=2.85 \pm 0.1$ mag.
However, the large errors below the turnoff make it difficult to 
clearly locate the $V$ magnitude of 
the bluest excursion of the isochrone which defines
the turnoff point.  However, if the 
red clumps of these clusters are forced to
coincide, the full dispersion in the V magnitudes
of the flat subgiant branches (a much better defined
point on the color-magnitude diagrams) is 0.2 mag.  
This is in good agreement with the 
dispersion in $\Delta V^{HB}_{TO}$.
Were we to use the extreme assumption that the youngest cluster
has [Fe/H]=--1 while the oldest has [Fe/H]=--1.66 we would find
the maximum permitted age spread to be  $7.5\pm 0.3 $ Gyr.  If the metallicities
of the clusters are nearly identical this spread reduces to $\pm 0.1$ Gyr.

 
If the cluster ridgelines are force-fit to the best coincidence possible,
the maximum dispersion at the turnoff is 0.1 $V$ mag.
Using the formula of
Buonanno et al. (1993) referred to in Section 3.0, we find that the maximum 0.1 
$V$ mag dispersion around the 
age of Kron 3 corresponds to $\pm 0.04$ dex in age.  For a mean age of 8 Gyr the
total duration of the burst is $\pm 0.7$ Gyr, or a full range of 1.4 Gyr.
Inspection of the Bertelli et al. (1994) isochrones (Figure 2) suggests
that the entire age spread is likely smaller than this, of order 1 Gyr.

\section{Are the Bursts Real?}

Our evidence for two bursts is based on the extraordinary agreement of the
cluster loci when forced to coincide.  This similarity reduces to
agreement in the $\Delta V^{HB}_{TO}$ index (age) and
in the red giant branch locus (metallicity).
Ages are usually determined by shifting isochrones according to 
distance and reddening and comparing the data to the turnoff in the isochrones.
This approach works poorly when distance and reddening is imprecisely known as 
is the case with the SMC clusters.  In order to test more rigorously
the narrowness of these 
bursts, we take 3 approaches toward age determination, all of which
derive from fitting the Bertelli et al. (1994) isochrones under
different assumptions:
  (1) We tie the 
horizontal branch of the isochrones to the cluster red HB.  This is the 
equivalent of using the $\Delta V^{HB}_{TO}$ method of age determination.
(2) We have shifted the isochrones according to the SMC distance modulus plus
total absorption based on 
the reddening derived for each cluster from the shifts in Table 6.  This
method assumes that the color shifts are due to reddening. (3) We derive
ages by placing the isochrones 
using the vdB91 values for the distance modulus (18.93 mag)
and reddening (0.04 mag).  Method (3) assumes that the color shifts
are perhaps due to an internal calibration error rather than measuring
actual differences in reddening.

Table 10 gives the summary of ages derived for the clusters using the 3 methods,
and Table 11 gives the age range for the young and old clusters, for 
[Fe/H] = $-1.71, -1.31,$ and $-0.71$.   For the young clusters, the best
fits for the isochrones were between the two high metallicities, while the
best fits for the old clusters occurred at the metal poor end, qualitatively
in agreement with Da Costa \& Hatzidimitriou (1998).  For the 2 Gyr burst
clusters, the full range of ages measured from
fits of the
isochrones is $1.0 < t_{Gyr} < 2.11$, however the 
smallest range was 1.6-2.1 Gyr, using the
HB fitting method.  Even allowing for a full variation in metallicity and age 
determination method, we find that the 3 young clusters formed
within a span of 0.5 Gyr in a burst
approximately 1.5 Gyr ago.

For the Kron 3 clusters, the full permitted age range of isochrone
fits, varying all the parameters, is 5-10 Gyr. 
 However, as Table 10 shows, the age range in each
entry is never larger than 2.6 Gyr and usually much smaller.  It is striking that 
when the HB fitting method is employed, the full range in permitted ages is 
$ 7.9 < t_{Gyr} < 8.9$ spanning the range $-1.71$ to $-1.31$ dex.  
The other two 
methods assume identical distance and reddening for all the old clusters;
as mentioned earlier, these assumptions lead to an age range that
is inconsistent with the excellent agreement in $\Delta V^{HB}_{TO}$
between the different CMDs. 
The outlier NGC 339 has a brighter clump than the other
clusters and may well lie 0.1 mag closer then the SMC while
NGC 416 is almost certainly reddened.  We conclude that the 8 Gyr old burst is also real.

It is instructive to consider in detail the implications of trying
to determine the ages and dispersion in ages using the standard
method of correcting the isochrones for the distance modulus and
reddening.
If we assume that the Kron 3 group lies at the SMC distance modulus and
that the shifts are due to reddening alone, the implied dispersion in age
is large.  The greatest contribution to a possible formal
age range would be a spread in metallicity: NGC 339 could be as
young as 3.1 Gyr with [Fe/H]=$-0.71$, and Kron 3 could be as old as
8.3 Gyr if its [Fe/H]=$-1.71$, a range of 5.3 Gyr.  However, the
lowest metallicity isochrones actually fit best while the more metal
rich isochrones are redder than the data.  Fitting with isochrones of 
the same (lowest) metallicity,
the age range permitted would be 2.6 Gyr.  We note that NGC 339 has the
brightest observed red clump of the entire sample, at $V=19.28$; at any
metallicity its age is found to be 1.5 Gyr younger than the other clusters
in the Kron 3 group.  We prefer the hypothesis that NGC 339 is $\approx 0.1$
mag closer than the bulk of the SMC.  If we were to assume that all the
clusters lie at the distance of the SMC then the derived age range 3-5
Gyr implies a 0.6-1.1 mag spread in $\Delta V^{HB}_{TO}$ using the
formula of Walker (1992) which would be clearly in disagreement with
the data.

In our opinion, the best measurement of relative ages for this
data set is the use of the $\Delta V^{HB}_{TO}$ method or the
equivalent method of fitting the Bertelli isochrones at the red
clump.  These approaches might be considered to be the
``minimum age spread'' solutions.  However, the assumption of
little age spread produces no additional observational
complications other than requiring that some combination of
distance and reddening produce a $10-20\%$ dispersion in the apparent
moduli of the clusters. 
Nor surprisingly, both of our methods based on the
red clump yield a dispersion in age of $10\%$ within
each burst.  

On the other hand, if we assume that all the clusters are at the
same distance and reddening and blindly fit the isochrones (or
alternatively, take our measured color-shifts as being due to
reddening alone) we find much larger age dispersions.  The
full range in age required would in turn demand a large
dispersion in $\Delta V^{HB}_{TO}$ that we do not observe,
and that would not be consistent with the isochrones themselves.

\subsection{Implied Dispersions in Distance Modulus and Reddening}

We have shown that the ridgelines of clusters in the NGC 411 
group are nearly identical.  Further, there is also 
little dispersion in the
reddening and distance moduli of the clusters in this group.
On the other hand, the Kron 3 clusters have similar ridgelines,
but have a 0.4 mag dispersion in the apparent brightness
of the red clump (see Figure 2). 
The discrepancy is especially evident for NGC 416,
The total dispersion in
$V$ magnitude for the Kron 3
clusters implied by both the shift method and the apparent $V$ magnitude of the 
red clump is $\approx 0.4$ mag.  The two most discrepant clusters are NGC 416 
and NGC 339.  These clusters have loci 0.1 mag redder than Kron 3,
and if due to reddening in the direction of these clusters, we must deredden 
both 
NGC 416 and NGC 339 and their 0.4 mag difference remains.  
If dereddened by 0.31 
mag in $V$ and 0.1 mag in $B-V$, the 
clump luminosity and locus of NGC 416 would 
be nearly identical to that of Kron 3.  Dereddening NGC 339 as required by its 
shift forces its red clump 0.43 mag brighter than Kron 3.  At a modulus of 
19.0, the total magnitude dispersion would correspond to a depth along the line 
of sight of $\approx 20\%\ \rm or \approx 10$ 
kpc $(\approx 10^o$ on the sky), 
if all of the dispersion were due to the depth of 
the pop II halo of the SMC (Westerlund 1997).  Because in fact NGC 339
accounts for nearly all of this dispersion, we do not take
this as an estimate of the 
depth of the SMC halo.

\section{The Oldest Clusters}

How do the clusters in the Kron 3 group compare with the 
oldest known SMC clusters?  Figure 3 compares the ridgeline of
Kron 3 with those of
NGC 121 and Lindsay 1 (Olszewski, Schommer, \& Aaronson 1987).
NGC 121 remains the oldest SMC cluster, followed by
Lindsay 1 (2 Gyr younger) and our Kron 3 group of clusters (4 Gyr younger than 
NGC 121).  Although the data sets for the old clusters include
ground-based photometry, we have redetermined the ages of these
clusters using our isochrones, in order to have a consistent
set of age measurements.
The 8 Gyr age determined above for the Kron 3 clusters is
consistent with the 12 Gyr age of NGC 121 (Stryker et al. 1985; Shara et al.
1998).  We derive an age of 10 Gyr for Lindsay 1, in good
agreement with Olszewski et al. (1987).
While we find evidence that the Kron 3 group is coeval, the old clusters NGC 121
and Lindsay 1 clearly differ in age from each other and from this group;
the oldest SMC clusters were not formed in a single short burst event.

\section{The Inferred Cluster Formation History}

It has long been accepted that the SMC has formed clusters continuously
during the past 12 Gyr.  Our revised formation history is illustrated
in Figure 4.  While the absolute age scale may change depending on the
choice of age indicator, the strikingly small duration of these bursts
is established via the empirical comparison of the new, homogeneous, HST
cluster color-magnitude
diagrams, which are identical within each group.  Referring back to Figure 2 we 
note the close agreement of the giant branches for all of the old SMC clusters, 
and we find no measurable dispersion in [Fe/H] within the clusters in the age 
range $8< t_9 < 12$ Gyr.

Figure illustrates our findings in comparison
with the roughly continuous cluster
formation history presented in Westerlund (1997) and other
reviews.  We also illustrate the burst history for the
most luminous $(M_V<-6)$ SMC clusters, which include all of those
imaged in our HST study.  Except for NGC 121, which is
clearly older than the 8 Gyr burst, these luminous clusters
all formed in either of the 2 or 8 Gyr bursts.
In the lowest panel of Figure 4 we include an additional
3 clusters from the literature, all of which have 
$13<V<14 (-6<M_V<-5)$: Lindsay 1 (Olszewski, Schommer, \& Aaronson 1987),
Lindsay 11 (Mould, Jensen, \& Da Costa  1992), and 
Lindsay 113 (Mould, Da Costa, \& Crawford 1984).  
Our ages for these
clusters, derived from $\Delta V_{TO}^{HB}$ (and using
Walker's 1992 formula), are $\approx 9$ Gyr for Lindsay 1
and 113, and $\approx 5$ Gyr for Lindsay 11.  These
ages are larger than those reported by the authors,
who fit the Yale isochrones.   
The age distribution in the lower histogram is 85\%
complete to $V\approx 14$.  
Although the sample sizes
are too small to draw any significant conclusion,
it is interesting that most of the luminous clusters 
formed in two discrete bursts. 

Given the apparent burst history and wide spatial separation
of these clusters, we believe that it is premature to consider
the notion of an age-metallicity relationship for the SMC
clusters.  The notion of chemical evolution contains the
implied assumption that enriched material from starbursts is
incorporated into later generations of stars.  Given
the widely spaced separation of the clusters and the
punctuated nature of the star formation, it is possible
that any enrichment of the interstellar medium occurring as
a consequence of the formation of these clusters affected
field stellar populations of the SMC and not the
abundances of later star clusters.  Note that the age-
metallicity relationship shows almost no change in the
abundances of clusters during the interval 5-10 Gyr
ago (Da Costa \& Hatzidimitriou  1998; Mighell et al. 1998)
It will be valuable to obtain detailed abundances
for a complete sample of SMC clusters, including the less luminous
clusters that we have not considered in this study.

Clusters can be destroyed as well as formed, however we note
that it is only the most luminous clusters that display the burst
history.  We have difficulty imagining a destruction mechanism
that would spare the less luminous clusters and leave two
neatly concentrated age groupings.

Ground-based color-magnitude diagrams yielded ages that 
were consistent with a roughly continuous distribution of
cluster ages that filled in the period of the LMC gap, the range of 2-14 Gyr.  
We agree that the SMC formed
clusters in that period, but we find evidence that the most
luminous clusters were formed in concentrated bursts.
As with the case of the field population of the LMC, we find
another example where large scale star formation has occurred
in bursts.  It is also interesting
that clusters spaced widely apart have such similar metal
abundances and ages.  These findings will require more complicated
chemical evolution models, and perhaps treating the field and
star clusters separately.

If we were able to observe an LMC-like galaxy at half the Hubble
time when the Kron 3 group was forming, we might observe four
widely separated bursts of star formation.  As many high redshift
galaxies have such a blotchy appearance, it is interesting to
speculate that we observe in these two groups of clusters the
fossil record of such events.

\acknowledgements

Support for this work was provided by NASA through grant number GO-5475
from the Space Telescope Science Institute, which is operated by AURA,
Inc., under NASA constract NAS5-2655.

\newpage

\figcaption{
The empirical ridgeline of NGC 411 is superposed on data from the
HST Planetary Camera for NGC 152, 419, and 411 in the SMC,
the ``NGC 411'' 2 Gyr burst group).  The lower right panel shows
the ridgelines (aligned using the red clump) and the Bertelli
et al. (1994) isochrones.  The best fit is for an age of 2 Gyr.}

\figcaption{
The empirical ridgeline of Kron 3 is superimposed on data from the HST
Planetary Camera for NGC 339, 361, 416, and Kron 3 in the 
SMC (the ``Kron 3'' 8 Gyr burst group).  Notice that the four clusters have 
nearly identical
principle
sequences: turnoff, red giant branch, and horizontal branch.
Within the quality of the data, there is no visible dispersion
in age, metallicity, or any other parameter between these clusters}

\figcaption{
{\it (Left Panel:)}The principle sequences of the 4
clusters in the 8 Gyr group 
are
aligned to match each other as closely 
as possible.  Also plotted
on this figure are the loci of Lindsay 1 and NGC 121, which are
illustrated because they are older.  Because the physical depth of the
SMC is large, we tie the CMD loci at the red clump to produce this
comparison.  The dispersion in main sequence turnoff loci of
the 8 Gyr group clusters is consistent
with being purely due to observational uncertainty.
{\it (Right Panel:)}  The Bertelli et al. (1994) isochrones are
fit to the Kron 3 ridgeline; the isochrones and the ridgeline
are aligned using the red clump.  The best fit isochrone corresponds
to 8 Gyr.}

\figcaption{
{\it (Upper Panel:)} The age distribution of old SMC clusters from Westerlund 
(1997) illustrates the continuous age distribution inferred from early
studies. {\it (Middle Panel:)} The age distribution derived from
our complete sample of clusters with $M_V<-6$; notice the two
very clear bursts.  {\it (Lower Panel:)}  Full sample of $>1$ Gyr old
SMC clusters with color-magnitude diagram; this sample includes
clusters as faint as $M_V=-5$.}

\end{document}


\begin{planotable}{lccc}
\tablewidth{300.0pt}
\tablecaption{Journal of Observations} 
\tablehead{
\colhead{Cluster} & \colhead{Date} & \colhead{Filter} & colhead{Exposure Time} \\ 
        & (dd/mm/yy) &   &    (Seconds)  }
\startdata
NGC 152 & 26/09/94 & F450W & 300.0  \\ 
        & 26/09/94 & F555W & 160.0  \\ 
NGC 339 & 07/04/94 & F450W & 400.0  \\
        & 07/04/94 & F555W & 200.0  \\
NGC 361 & 03/04/94 & F450W & 300.0  \\
        & 03/04/94 & F555W & 160.0  \\
NGC 411 & 25/05/94 & F450W & 400.0  \\
        & 25/05/94 & F555W & 200.0  \\
NGC 416 & 06/02/94 & F450W & 400.0  \\
        & 06/02/94 & F555W & 200.0  \\
NGC 419 & 08/04/94 & F450W & 400.0  \\
        & 08/04/94 & F555W & 200.0  \\
KRON 3  & 27/05/94 & F450W & 600.0  \\
        & 27/05/94 & F555W & 300.0  \\
\strut \\ \hline
\end{planotable}

\begin{planotable}{lcc}
\tablewidth{200.0pt}
\tablecaption{Aperature Corrections} 
\tablehead{
\colhead{Cluster} & \colhead{F555W}     & \colhead{F450W}  \\ 
        &   (mag)   &    (mag)    } 
\startdata
NGC 152 & 0.1849 & 0.2213 \\
NGC 339 & 0.1704 & 0.1775 \\
NGC 361 & 0.1729 & 0.1974 \\
NGC 411 & 0.2014 & 0.2040 \\
NGC 416 & 0.2033 & 0.2242 \\
NGC 419 & 0.1927 & 0.1833 \\
KRON 3  & 0.1802 & 0.2021 \\
\strut \\ \hline
\end{planotable}

\begin{planotable}{lcc}
\tablewidth{200.0pt}
\tablecaption{Pre/Post Cool Down Corrections} 
\tablehead{
\colhead{Cluster} & \colhead{F555W}     & \colhead{F450W}  \\ 
        &   (mag)   &    (mag)    } 
\startdata
NGC 152 &  0.000 &  0.000 \\
NGC 339 & -0.062 & -0.078 \\
NGC 361 & -0.062 & -0.077 \\
NGC 411 &  0.000 &  0.000 \\
NGC 416 & -0.062 & -0.080 \\
NGC 419 & -0.062 & -0.078 \\
KRON 3  &  0.000 &  0.000 \\
\strut \\ \hline
\end{planotable}

\begin{planotable}{lcc}
\tablewidth{200.0pt}
\tablecaption{Contamination Corrections} 
\tablehead{
\colhead{Cluster} & \colhead{F555W}     & \colhead{F450W}  \\ 
        &   (mag)   &    (mag)    } 
\startdata
NGC 152 & -0.001 & -0.001 \\
NGC 339 & -0.007 & -0.004 \\
NGC 361 & -0.005 & -0.003 \\
NGC 411 &  0.000 &  0.000 \\
NGC 416 & -0.007 & -0.004 \\
NGC 419 & -0.014 & -0.008 \\
KRON 3  & -0.002 & -0.001 \\
\strut \\ \hline
\end{planotable}

\begin{planotable}{lcc}
\tablewidth{200.0pt}
\tablecaption{Cluster Offsets for fit} 
\tablehead{
\colhead{Cluster} & \colhead{$\delta$V} & \colhead{$\delta$B-V}  \\ 
        &   (mag)   &    (mag)     }
\startdata
NGC 121   & -0.22 & -0.11 \\ 
NGC 152   &  0.00 &  0.00 \\ 
NGC 339   & +0.13 & -0.09 \\ 
NGC 361   & +0.02 & -0.06 \\ 
NGC 411   & +0.20 & -0.06 \\ 
NGC 416   & -0.28 & -0.10 \\ 
NGC 419   & +0.16 & -0.02 \\ 
KRON 3    &  0.00 &  0.00 \\ 
LINDSAY 1 & +0.11 & -0.16 \\ 
\strut \\ \hline
\end{planotable}

\begin{planotable}{cc}
\tablewidth{100.0pt}
\tablecaption{Kron 3 Ridgeline} 
\tablehead{
\colhead{B-V} & \colhead{V}  \\ }
\startdata
        0.398     &    23.873  \\
        0.377     &    23.702  \\
        0.357     &    23.470  \\
        0.335     &    23.232  \\
        0.307     &    22.945  \\
        0.293     &    22.713  \\
        0.288     &    22.530  \\
        0.281     &    22.353  \\
        0.281     &    22.191  \\
        0.282     &    22.061  \\
        0.293     &    21.983  \\
        0.318     &    21.918  \\
        0.356     &    21.868  \\
        0.398     &    21.843  \\
        0.440     &    21.818  \\
        0.486     &    21.818  \\
        0.516     &    21.823  \\
        0.533     &    21.806  \\
        0.552     &    21.781  \\
        0.570     &    21.735  \\
        0.580     &    21.671  \\
        0.592     &    21.574  \\
        0.606     &    21.409  \\
        0.618     &    21.160  \\
        0.625     &    20.953  \\
        0.640     &    20.488  \\
        0.664     &    20.214  \\
        0.687     &    19.965  \\
        0.720     &    19.704  \\
        0.748     &    19.443  \\
        0.780     &    19.269  \\
        0.808     &    19.045  \\
        0.874     &    18.647  \\
        0.976     &    18.100  \\
        1.112     &    17.541  \\
        1.341     &    16.969  \\
        1.485     &    16.857  \\
        1.653     &    16.819  \\
\strut \\ \hline
\end{planotable}

\begin{planotable}{cc}
\tablewidth{100.0pt}
\tablecaption{NGC 419 Ridgeline} 
\tablehead{
\colhead{B-V} & \colhead{V}  \\ }
\startdata
        0.486    &     23.372  \\
        0.444    &     23.111  \\
        0.365    &     22.738  \\
        0.318    &     22.353  \\
        0.285    &     22.092  \\
        0.271    &     21.967  \\
        0.239    &     21.793  \\
        0.220    &     21.582  \\
        0.206    &     21.420  \\
        0.187    &     21.246  \\
        0.183    &     21.097  \\
        0.183    &     20.997  \\
        0.173    &     20.898  \\
        0.173    &     20.774  \\
        0.173    &     20.660  \\
        0.169    &     20.528  \\
        0.184    &     20.411  \\
        0.229    &     20.312  \\
        0.293    &     20.254  \\
        0.345    &     20.254  \\
        0.416    &     20.295  \\
        0.453    &     20.362  \\
        0.506    &     20.411  \\
        0.565    &     20.486  \\
        0.606    &     20.519  \\
        0.666    &     20.561  \\
        0.718    &     20.519  \\
        0.734    &     20.363  \\
        0.748    &     20.251  \\
        0.757    &     20.139  \\
        0.775    &     19.964  \\
        0.797    &     19.723  \\
        0.816    &     19.574  \\
        0.836    &     19.369  \\
        0.869    &     19.095  \\
        0.898    &     18.903  \\
        0.930    &     18.697  \\
        0.981    &     18.336  \\
        1.004    &     18.138  \\
        1.032    &     17.988  \\
        1.088    &     17.653  \\
        1.144    &     17.404  \\
        1.201    &     17.205  \\
        1.243    &     17.056  \\
        1.275    &     16.956  \\
        1.294    &     16.894  \\
\strut \\ \hline
\end{planotable}

\begin{planotable}{cc}
\tablewidth{100.0pt}
\tablecaption{KRON 3 HB Ridgeline} 
\tablehead{
\colhead{B-V} & \colhead{V} \\ }
\startdata
        0.664    &   19.493 \\
        0.640    &   19.493 \\
        0.612    &   19.493 \\
        0.575    &   19.480 \\
        0.542    &   19.493 \\
        0.505    &   19.480 \\
        0.472    &   19.480 \\
\strut \\ \hline
\end{planotable}

\begin{planotable}{cc}
\tablewidth{100.0pt}
\tablecaption{NGC 419 HB Ridgeline} 
\tablehead{
\colhead{B-V} & \colhead{V} \\ }
\startdata
        0.743    &     19.443  \\
        0.724    &     19.456  \\
        0.706    &     19.443  \\
        0.682    &     19.456  \\
        0.659    &     19.468  \\
        0.626    &     19.456  \\
        0.594    &     19.468  \\
        0.557    &     19.468  \\
        0.519    &     19.468  \\
\strut \\ \hline
\end{planotable}

\begin{planotable}{llccc}
\tablewidth{400.0pt}
\tablecaption{Cluster Age Determinations From Isochrone Fits} 
\tablehead{
\colhead{Cluster} & \colhead{Method} & \colhead{[Fe/H]=-1.71} & \colhead{[Fe/H]=-1.31} & \colhead{[Fe/H]=-0.71} \\ 
    &    &   t (Gyr) & t (Gyr) & t (Gyr) }
\startdata
NGC 152 & SMC mod only  & 1.8 - 2.5  & 1.6 - 2.2  & 1.3 - 2.0 \\
        & mod + shift   & 1.8 - 2.5  & 2.0 - 2.5  & 1.3 - 1.8 \\
        & forced HB fit & 1.0 - 1.4  & 1.6 - 2.2  & 1.3 - 1.8 \\
NGC 411 & SMC mod only  & 1.3 - 2.0  & 1.4 - 2.2  & 1.0 - 1.6 \\
        & mod + shift   & 1.3 - 2.0  & 1.3 - 2.0  & 1.0 - 1.6 \\
        & forced HB fit & 0.8 - 1.3  & 1.3 - 2.0  & 1.3 - 1.8 \\
NGC 419 & SMC mod only  & 1.3 - 2.2  & 1.4 - 2.2  & 1.0 - 1.8 \\
        & mod + shift   & 1.6 - 2.2  & 1.8 - 2.2  & 1.0 - 1.6 \\
        & forced HB fit & 0.8 - 1.3  & 1.3 - 2.0  & 1.3 - 1.8 \\
Kron 3  & SMC mod only  & 7.1 - 8.9  & 5.6 - 7.9  & 3.5 - 5.6 \\
        & mod + shift   & 7.1 - 8.9  & 5.6 - 7.1  & 4.5 - 5.6 \\
        & forced HB fit & 7.1 - 8.9  & 7.1 - 8.9  & 6.3 - 7.1 \\
NGC 339 & SMC mod only  & 6.3 - 10.0 & 5.0 - 7.9  & 3.5 - 5.6 \\
        & mod + shift   & 5.0 - 6.3  & 4.0 - 5.0  & 2.8 - 3.5 \\
        & forced HB fit & 7.1 - 8.9  & 7.1 - 8.9  & 6.3 - 7.1 \\
NGC 361 & SMC mod only  & 7.1 - 10.0 & 6.3 - 8.9  & 4.0 - 6.3 \\
        & mod + shift   & 7.1 - 8.9  & 5.0 - 6.3  & 3.5 - 4.5 \\
        & forced HB fit & 7.9 - 10.0 & 7.9 - 8.9  & 7.1 - 8.9 \\
NGC 416 & SMC mod only  & 8.9 - 11.2 & 7.1 - 11.2 & 4.0 - 7.1 \\
        & mod + shift   & 7.1 - 8.9  & 5.6 - 7.1  & 4.0 - 5.0 \\
        & forced HB fit & 7.9 - 8.9  & 7.1 - 8.9  & 5.6 - 7.1 \\
\strut \\ \hline
\end{planotable}

\begin{planotable}{llccc}
\tablewidth{400.0pt}
\tablecaption{Cluster Age Determinations From Isochrone Fits} 
\tablehead{
\colhead{Cluster} & \colhead{Method} & \colhead{[Fe/H]=-1.71} & \colhead{[Fe/H]=-1.31} & \colhead{[Fe/H]=-0.71} \\ 
    &    & t (Gyr) & t (Gyr) & t (Gyr) }
\startdata
NGC 411 & SMC mod only  & 1.58 - 2.11 & 1.78 - 2.00 & 1.26 - 1.58 \\
        & mod + shift   & 1.58 - 2.11 & 1.58 - 2.11 & 1.26 - 1.50 \\
        & forced HB fit & 1.00 - 1.19 & 1.58 - 2.11 & 1.26 - 1.50 \\
Kron 3  & SMC mod only  & 7.94 - 10.0 & 6.31 - 8.91 & 4.47 - 6.31 \\
        & mod + shift   & 5.62 - 7.94 & 4.47 - 7.24 & 3.16 - 5.62 \\
        & forced HB fit & 7.94 - 8.91 & 7.94 - 8.41 & 6.31 - 7.94 \\
\strut \\ \hline
\end{planotable}